\documentclass[a4paper]{jpconf}
\usepackage[dvips]{graphicx}
\usepackage{latexsym} 

\newcommand{\beqn}{\begin{eqnarray}}
\newcommand{\eeqn}{\end{eqnarray}}

\begin{document}
\title{Testing Superdeterministic Conspiracy}

\author{Sabine Hossenfelder}

\address{Nordita, KTH Royal Institute of Technology and Stockholm University\\
Roslagstullsbacken 23, SE-106 91 Stockholm, Sweden}

\ead{hossi@nordita.org}

\begin{abstract}
Tests of Bell's theorem rule out local hidden variables theories. But any theorem is only as good as the assumptions that go into it, and one of these assumptions is that the experimenter can freely chose the detector settings.  Without this assumption, one enters the realm of superdeterministic hidden variables theories and can no longer use Bell's theorem as
a criterion. One can like or not like such superdeterministic hidden variables theories and their inevitable nonlocality, the real question is how
one can test them. Here, we propose a possible experiment that could reveal superdeterminism.
\end{abstract}

\section{Introduction}

For any locally causal theory, the attempt to explain quantum effects by use of randomness
induced through hidden variables can be shown to be in disagreement with experiment 
via Bell's theorem \cite{Bell:1964kc}, or its generalization respectively \cite{Clauser:1969ny}. 
Throughout the last decades, experiments 
have established that the hidden variables theories
for which Bell's theorem applies are not realized in nature \cite{Aspect:1982fx,Weihs:1998gy,Tittel:1998ja,Rowe,Moehring,Ansmann,Hasegawa} . There are various loopholes
in the conclusions that can be drawn from these experiments and not all of these
loopholes have yet been satisfactory closed (for a recent summary see e.g. \cite{loop}). But even so,
local hidden variables are strongly disfavored. 

However, Bell's theorem uses the
assumption that one has the freedom to choose the detector's settings without modifying the 
prepared state that one wishes to measure. Without this freedom the
conclusion of Bell's theorem does not apply.  Superdeterminism requires a non-local
correlation between the prepared state and the detector which seems unappealing. 
But the question whether the 
time evolution of our universe is fundamentally deterministic
or not is too important to leave it to taste -- it is an hypothesis that must be tested
experimentally. 

\section{A Proposal for an Experiment}

The essential difference between standard quantum mechanics and superdeterministic hidden variables theories is that in the former case two identically prepared states can give two different measurement outcomes, while in the latter case that is not possible. Unfortunately, `identically prepared' includes the hidden variables, and it is difficult to identically prepare something one cannot measure. That is after all the reason why it looks indeterministic.

\begin{figure}
\begin{center}
\includegraphics[width=8cm]{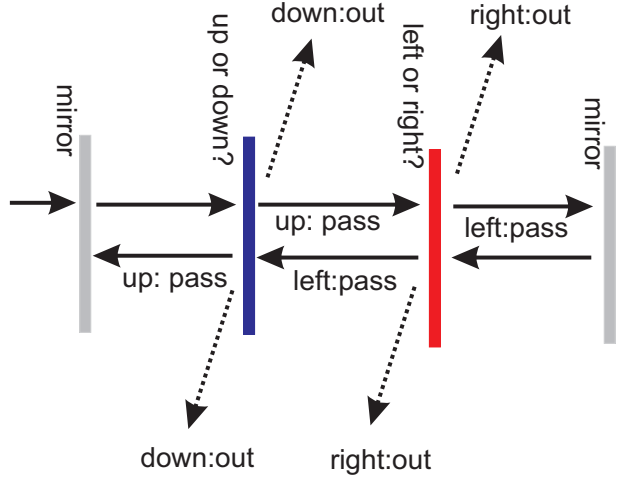}
\end{center}
\caption{\label{fig1} Sketch of proposed experiment. The prepared state is bounced between two mirrors. Two 
non-commuting observables (for example the up-down and left-right component of spin) are measured alternatingly.
Only one combination passes. In standard quantum mechanics, the probability for the particle to pass decreases
exponentially with time. In a superdeterministic theory with hidden variables, a particle that passes once will in principle
continue to pass indefinitely. In practice, it will pass until noise destroys the autocorrelation of the time-series.}
\end{figure}

To circumvent this problem, instead of trying to prepare identical states one can make repeated measurements on the same state. For that, one takes two non-commuting variables (for example the spin or polarization in two different directions) and measures them alternately. In standard quantum mechanics the measurement outcomes will be non-correlated; the measurement of one observable prevents the state from being in an eigenstate of the other observable. In a superdeterministic hidden variables theory, the observables will be correlated - provided one can make a case that the hidden variables do not change in between the measurements. Figure \ref{fig1} depicts an example for an experimental setup.

The provision that the hidden variables do not change is the reason why the test is only `almost' model independent because it brings
in an assumption about the origin of the hidden variables which is necessary to make the hypothesis testable at all. To arrive at
concrete estimates that are explained in more detail in \cite{Hossenfelder:2011ct} , we thus assume that the hidden variables are due to the environment (the experimental setup) down to the relevant scales of the interactions taking place. That means we take on
an effective field theory perspective and decouple all scales below the resolution of the experiment. As so often when it comes to the foundations of quantum mechanics, 
the noise is the main problem in this experimental setup. In practice, we will have to make the system small and cool and measure quickly enough so we have a chance to see the correlation between subsequent measurements.  

To estimate whether it is possible to measure the correlations, let us denote one of the two non-commuting observables
with $O$ and label with $\kappa$ the times at with $O$ is measured. We are then interested in the
autocorrelation time of the time-series of measurements of $O$. To simplify matters, we can look
at the correlation of subsequent measurements with the first measurement:

\beqn
{\mathrm{Corr}}_\kappa = \frac{E(\phi, O, t=0 \wedge \phi, O, t=\kappa )}{E(\phi, O^2)}~,
\eeqn
where $E$ denotes the expectation value.

For the autocorrelation we make the usual ansatz of exponential decay
\beqn
{\mathrm{Corr}}_\kappa = \exp \left( - \kappa/\tau \right) ~,
\eeqn
where $\tau$ is some timescale, the autocorrelation time, that encodes the rate of change of the experimental
setup -- the noise -- and thus the environmentally induced hidden variables. Here, we implicitly assumed that the
divergence of the system from a perfect repetition happens by incremental disturbances that are described 
by a homogeneous Poisson-process, i.e. the changes that lead to the decay of the autocorrelation 
are statistically independent and uniform in time. The larger the experiment and the shorter the typical timescales 
on which the relevant degrees of freedom change, 
the smaller $\tau$ and the faster the correlation will vanish. In the limit $\tau \to 0$,
one reproduces standard quantum mechanics. 

The question of whether one can test superdeterministic hidden variables theories then
comes down to the question of whether we can find a setting in which the autocorrelation time is
large enough to be measureable.

Consider as a concrete example a photodetector constituted of $N$ atoms. Infalling
photons excite an electron into the conduction band with an energy gap of $\Delta E$. 
At a temperature $T$ one single atom has a probability of $\exp(-\Delta E/T)$
to become excited by thermal motion, and it remains so for the average 
electron-hole recombination time $\tau_{\mathrm r}$ which is typically of the order of
a nanosecond. The 
time $\tilde \tau$ for $N$ atoms to undergo a statistical change is then
\beqn
\tilde \tau \approx \frac{\exp(\Delta E/T)~ }{N} \tau_{\mathrm r} ~.
\eeqn
There are many other sorts of noise, but this is the one with the highest
frequency, and thus the one relevant for the decay time.

A microscopically small detector with an extension of some $\mu$m would be constituted of approximately
$N \approx 10^{15}$ atoms. A semi-conductor with a fairly large band gap of $\Delta E \approx 1$~eV
at $T \sim 300$K then has $\tilde \tau \approx 10^{-6}$s. This autocorrelation time is feasible to
measure and is also considerably larger than the time it would take a photon to bounce back between
the mirrors in such a setup. We therefore have reason to be cautiously optimistic that such an experimental
test can be realized with today's technological possibilities.

\section{History}

Interestingly, in a footnote of a 1970 paper  \cite{Wigner:1976ga}, Eugene Wigner mentions Von Neumann discussing exactly this type of experiment: 
\begin{quote}``Von Neumann often discussed the measurement of the spin component of a spin-1/2 particle in various directions. Clearly, the possibilities for the two possible outcomes of a single such measurement can be easily accounted for by hidden variables [...] However, Von Neumann felt that this is not the case for many consecutive measurements of the spin component in various different directions. The outcome of the ﬁrst such measurement restricts the range of values which the hidden parameters must have had before that ﬁrst measurement was undertaken. The restriction will be present also after the measurement so that the probability distribution of the hidden variables characterizing the spin will be different for particles for which the measurement gave a positive result from that of the particles for which the measurement gave a negative result. The range of the hidden variables will be further restricted in the particles for which a second measurement of the spin component, in a different direction, also gave a positive result...''
\end{quote}

A longer discussion with Schr\"odinger followed this proposal. The conclusions could be summarized with saying that the experiment cannot test generic superdeterminism, but only certain types. If we think about it for a moment, this is unquestionably
true because we can never rule out generic superdeterminism anyway: If time evolution was fundamentally indeterministic, one could
always postulate some nonlocal hidden variables that remain unmeasurable. Such an hypothesis is however scientifically uninteresting because it cannot be falsified. To make superdeterminism scientifically interesting, one has to make additional
assumptions about the origin of the hidden variables. Only then can one test it.

\section{Conclusions}

We have here proposed an experiment to test superdeterministic theories with hidden variables that
are environmentally induced and estimated the prospects to realized this test. We have argued that
it seems possible to test this type of theory with presently existing technology. Such an experiment
could reveal that the apparent indeterminism of quantum mechanics is not fundamental and the
collapse of the wave-function is not fundamentally probabilistic.

\end{document}